\documentclass{webofc}
\usepackage[varg]{txfonts}   % Web of Conferences font
\usepackage{hyperref}
\usepackage{url}
\hypersetup{colorlinks=true,citecolor=blue,urlcolor=blue,linkcolor=blue}
\usepackage{siunitx}
\usepackage{graphicx}
\usepackage{subcaption}
\usepackage{lineno}
\usepackage{xcolor}

\makeatletter
\newcommand\MyAutoefPhrasecolorGroup[1]{%
  \color@begingroup\color{MyCurrentcolor}#1\endgroup
}%
\def\HyRef@testreftype#1.#2\\{%
 \colorlet{MyCurrentcolor}{.}%
 \ltx@IfUndefined{#1autorefname}{%
   \ltx@IfUndefined{#1name}{%
     \HyRef@StripStar#1\\*\\\@nil{#1}%
     \ltx@IfUndefined{\HyRef@name autorefname}{%
       \ltx@IfUndefined{\HyRef@name name}{%
         \def\HyRef@currentHtag{}%
         \Hy@Warning{No autoref name for `#1'}%
       }{%
         \edef\HyRef@currentHtag{%
           \noexpand\MyAutoefPhrasecolorGroup{%
             \expandafter\noexpand\csname\HyRef@name name\endcsname
           }%
           \noexpand~%
         }%
       }%
     }{%
       \edef\HyRef@currentHtag{%
         \noexpand\MyAutoefPhrasecolorGroup{%
           \expandafter\noexpand
           \csname\HyRef@name autorefname\endcsname
         }%
         \noexpand~%
       }%
     }%
   }{%
     \edef\HyRef@currentHtag{%
       \noexpand\MyAutoefPhrasecolorGroup{%
         \expandafter\noexpand\csname#1name\endcsname
       }%
       \noexpand~%
     }%
   }%
 }{%
   \edef\HyRef@currentHtag{%
     \noexpand\MyAutoefPhrasecolorGroup{%
       \expandafter\noexpand\csname#1autorefname\endcsname
     }%
     \noexpand~%
   }%
 }%
}%
\makeatother

\begin{document}
\def\scaler{0.75}
\renewcommand{\figureautorefname}{Fig.}

\title{IceCube Upgrade status and perspectives}

\author{\firstname{Anna} \lastname{Eimer}\inst{1,2}\fnsep\thanks{\email{anna.eimer@fau.de}} for the IceCube collaboration \footnote{\protect\url{http://icecube.wisc.edu}}
        % etc.
}

\institute{Erlangen Centre for Astroparticle Physics 
\and
           Friedrich-Alexander-Universität Erlangen-Nürnberg
          }

\abstract{The IceCube Neutrino Observatory instruments one cubic kilometer of deep-glacial ice between \SI{1450}{m} and \SI{2450}{m} below the surface at the geographic South Pole to detect neutrinos via Cherenkov radiation of relativistic charged particles produced in their interactions. This detector is responsible for a number of key observations in neutrino astrophysics, which include the discovery of a high-energy astrophysical neutrino flux and, more recently, the galactic plane.

During the austral summer of \num{2025}/\num{26}, five new strings equipped with new photosensor designs were deployed as a dense infill in the middle of the existing detector. The science goals of this detector are twofold: Firstly, given the higher photocathode density, an improved atmospheric neutrino event selection and reconstruction at a few \si{GeV} can be achieved for enhanced capabilities to study neutrino oscillations. Secondly, novel calibration devices will improve the knowledge of the optical properties of the glacial ice and the detector response. These new calibration results will be applied to archival IceCube data, improving angular and spatial resolution of all detected astrophysical neutrino events. The IceCube Upgrade also serves as a first step towards the next-generation neutrino telescope at the South Pole, called IceCube-Gen2.
}
\maketitle
\section{Introduction}\label{intro}
The IceCube Neutrino Observatory is a cubic-kilometer neutrino detector installed in the glacial ice at the geographic South Pole \cite{IceCube_instrumentation_and_online_system}. It detects Cherenkov radiation emitted by secondary particles produced in neutrino interactions in the surrounding ice or the nearby bedrock.
%The first configuration of the IceCube Neutrino Observatory \cite{IceCube_instrumentation_and_online_system} was completed in \num{2011}. It detects neutrinos via the Cherenkov effect. Neutrinos interact with the glacial ice at the geographical South Pole and produce charged secondary particles. These secondaries, for example muons or electrons, travel faster than the speed of light in the ice and emit Cherenkov radiation. The emitted light travels through the glacial ice to the photosensors, called digital optical modules (short DOMs) containing a photomultiplier tube (PMT) for the detection.

The original in-ice array of IceCube, completed in \num{2011}, consists of \num{86} instrumented cables, called strings. Each string features \num{60} photosensors, called digital optical modules (DOMs), which are located at depths between \SI{1450}{m} and \SI{2450}{m} in the ice. The string separation is mostly \SI{125}{m} with  DOMs spaced \SI{17}{m} apart vertically. This leads to detectable energies in the \si{TeV} to \si{PeV} range. IceCube also features a low-energy infill called DeepCore \cite{IceCube:2011ucd}. Here the string spacing amounts to \SI{72}{m}, and the DOM separation is \SI{7}{m}. This denser spacing lowers the energy threshold to about \SI{10}{GeV}. All DOMs are equipped with a single downward-facing \SI{10}{inch} photomultiplier tube (PMT) as a photosensor.  

In this detector configuration, several discoveries have been made over the years. A diffuse astrophysical neutrino flux was discovered in \cite{3year_diffuse_flux}, and the most recent studies revealed a spectral break at \SI{10}{TeV} \cite{spectral_break}. More recently, in \num{2023}, the neutrino emission from the galactic plane \cite{galactic_plane} was established and now confirmed with \num{5}$\sigma$ \cite{galactic_plane_5sigma}. Also, several extragalactic neutrino source candidates have been observed, including $\gamma$-ray blazars \cite{txs0506} and Seyfert galaxies \cite{ngc1068}. At this point, these high-energy observations are mostly limited by statistics and the knowledge about the detection medium, the glacial ice. The first measurement of the galactic plane, published in 2023, was partially made possible by an improved angular resolution resulting from updates to the ice optical modeling \cite{improved_shower_modeling}.

\section{The IceCube Upgrade}\label{status}

In the austral summer of \num{2025}/\num{2026}, % fifteen years after finishing the construction of IceCube, 
five new strings have been added to the detector, forming the so-called  IceCube Upgrade \cite{upgrade}. The Upgrade aims to lower the energy threshold to the few \si{GeV} regime in order to study neutrino oscillation with better precision. Other goals are to explore new photosensor designs and to further improve the detector calibration by a detailed study of the optical properties of the old glacial ice and the refrozen drill holes.

\subsection{String layout}\label{string_layout}

\begin{figure}[htbp]
\centering
\includegraphics[width=\linewidth,clip]{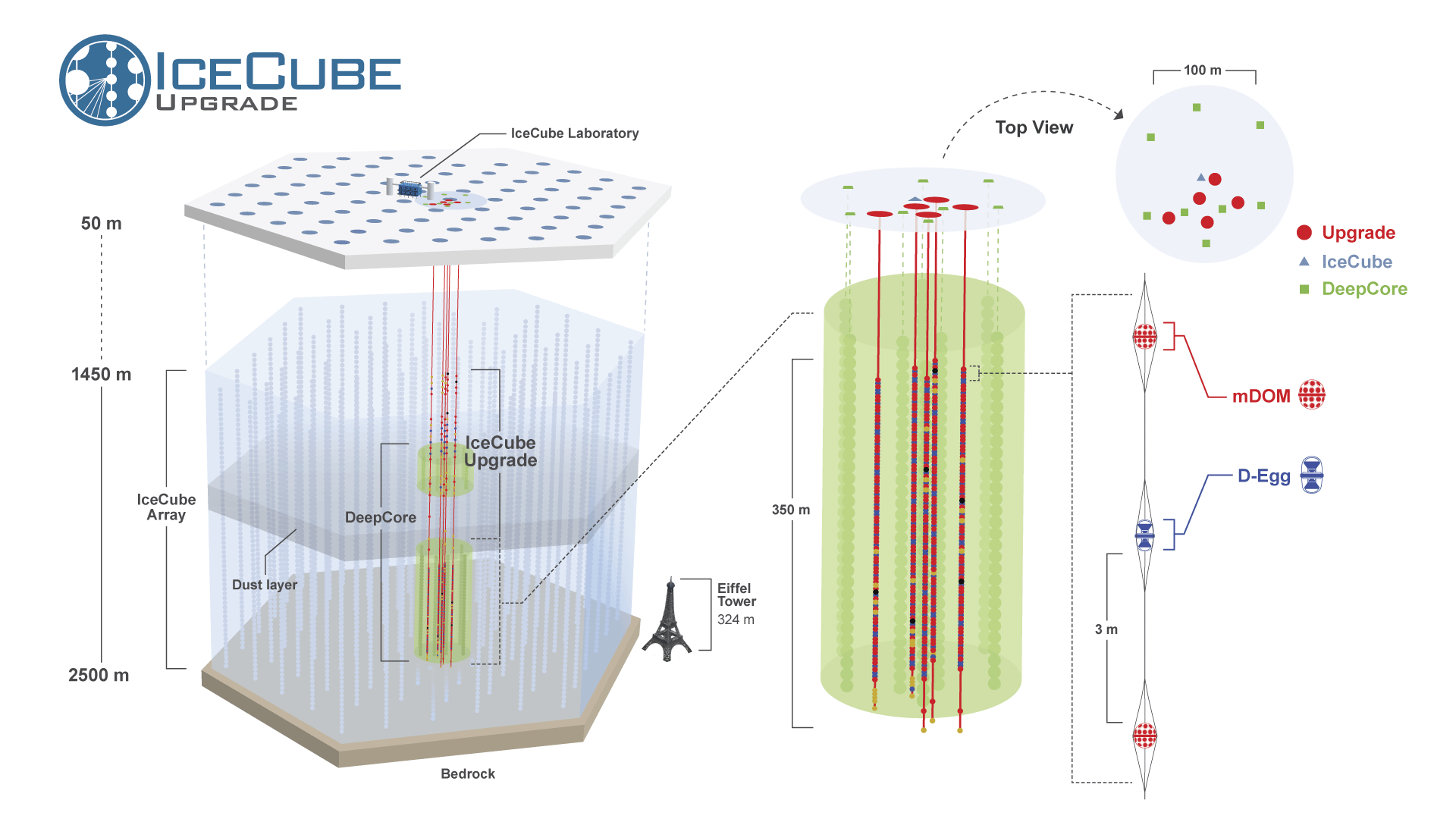}
\caption{Detector layout. Left: The detector layout of IceCube. The green area marks the DeepCore region and the red strings show the position of the IceCube Upgrade strings. Right: A zoomed-in version of the DeepCore region showing the layout of the IceCube Upgrade strings and a top view of the region.}
\label{fig:detector_layout}
\end{figure}

To lower the energy threshold, an average string spacing of \SI{20}{m} and a vertical DOM separation of \SI{3}{m} were chosen. The final string layout is shown in \autoref{fig:detector_layout}. It is fully contained within the DeepCore region. %Six new strings were deployed within one season. The first newly deployed string lost communication, presumably due to a cable break at \SI{700}{m}. 
The \num{5} commissioned strings work well and are expected to start taking science data with the rest of the detector in fall \num{2026}.

\subsection{Deployed photosensor modules}\label{photosensor_modules}

\begin{figure}[htbp]
  \centering
  %--------- linker Block: zwei Zeilen ----------
  \begin{minipage}[t]{0.77\textwidth}
    \vspace{0pt}% Oberkanten der beiden Minipages ausrichten
    % Zeile 1: zwei Querformat-Bilder
    \begin{subfigure}[b]{0.48\linewidth}
    \centering
      \includegraphics[width=\scaler\linewidth]{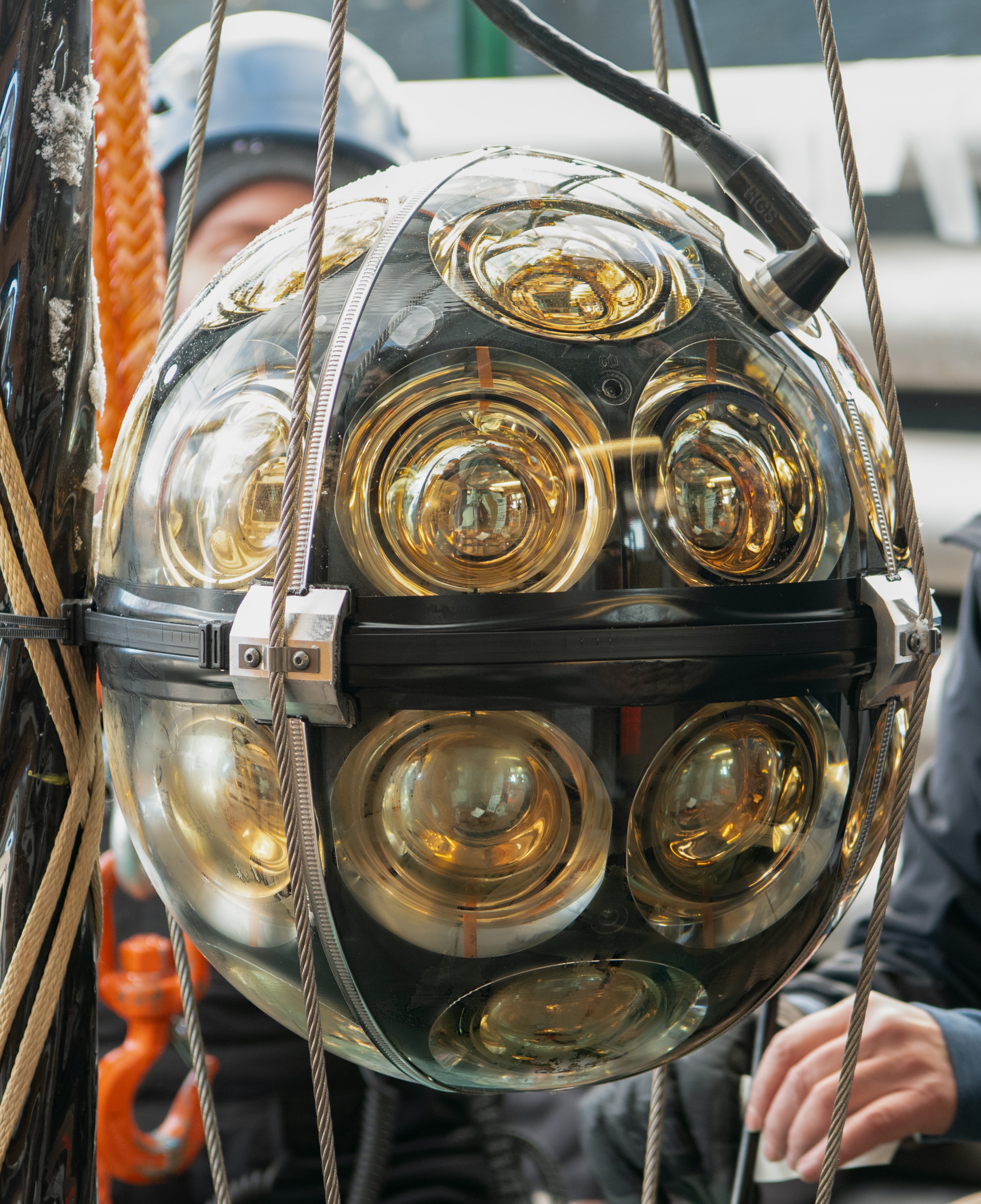}
      \caption{mDOM}
        \label{fig:mdom}
    \end{subfigure}%\hfill
    \begin{subfigure}[b]{0.4032\linewidth}
        \centering
      \includegraphics[width=\scaler\linewidth]{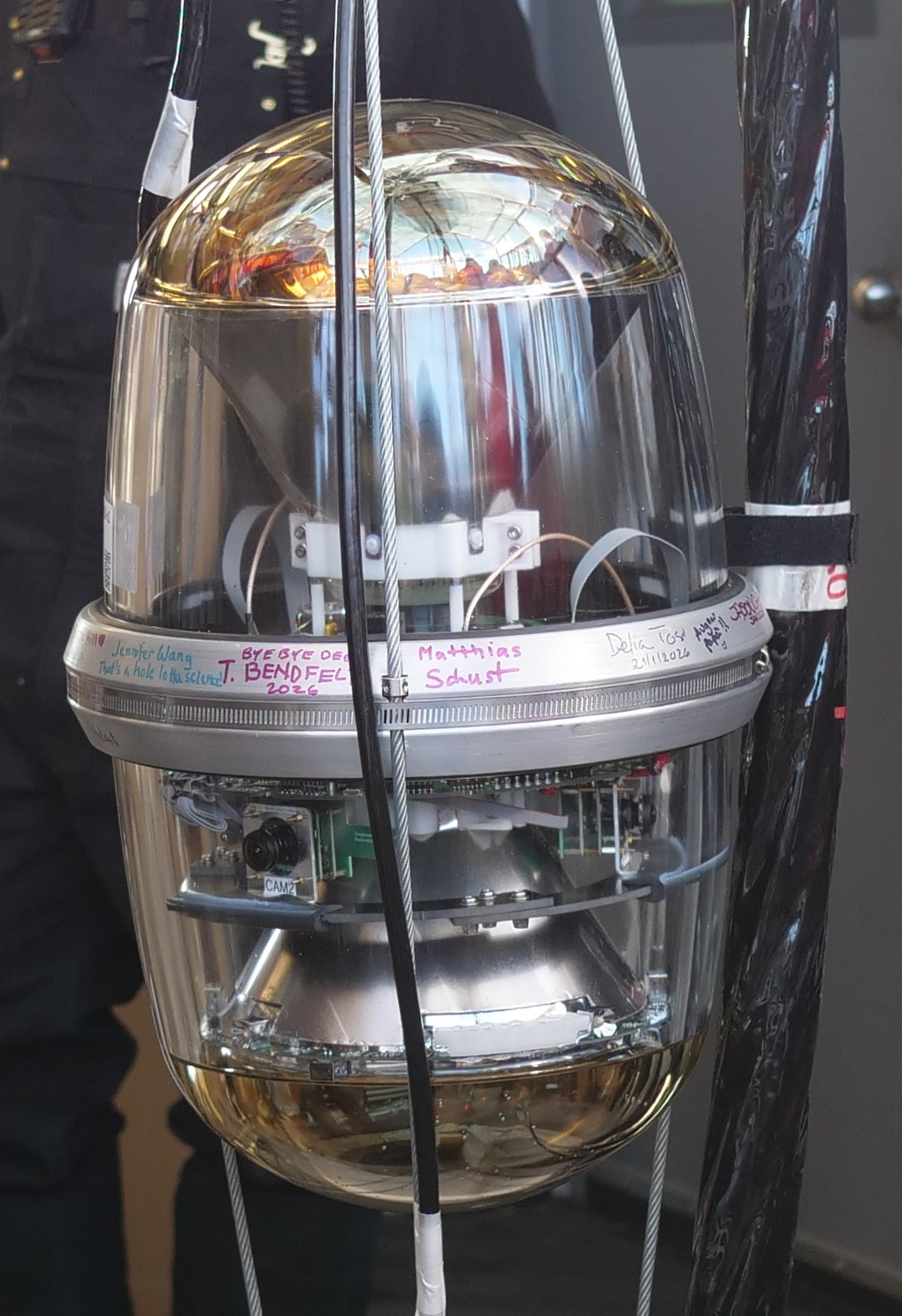}
      \caption{D-Egg}
        \label{fig:degg}
    \end{subfigure}

    \vspace{1ex}

    % Zeile 2: drei Querformat-Bilder
    \begin{subfigure}[b]{0.34\linewidth}
    \centering
      \includegraphics[width=\scaler\linewidth]{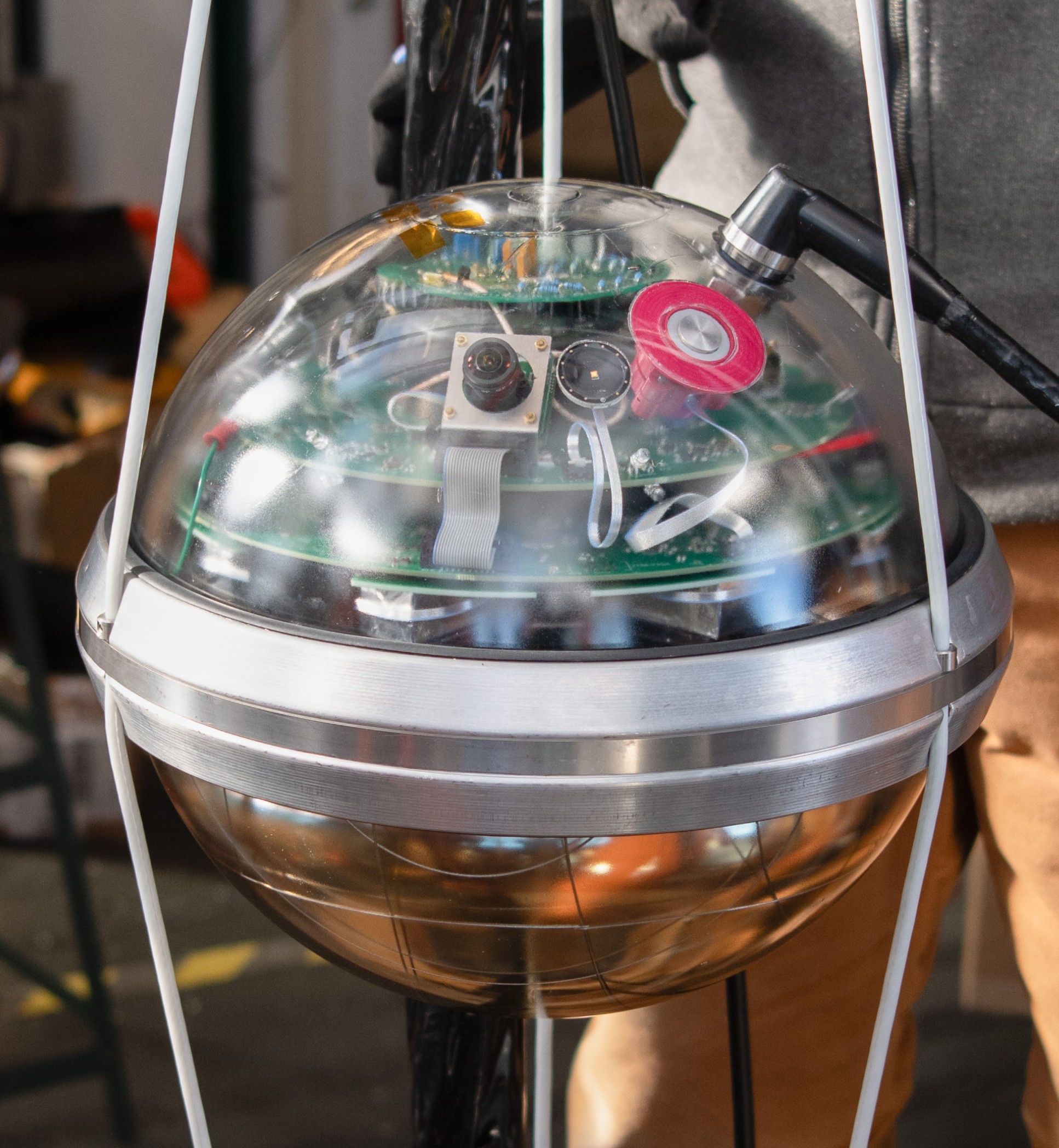}
      \caption{retro fitted Gen1 DOM}
        \label{fig:pdom}
    \end{subfigure}%\hfill
    \begin{subfigure}[b]{0.33\linewidth}
    \centering
      \includegraphics[width=\scaler\linewidth]{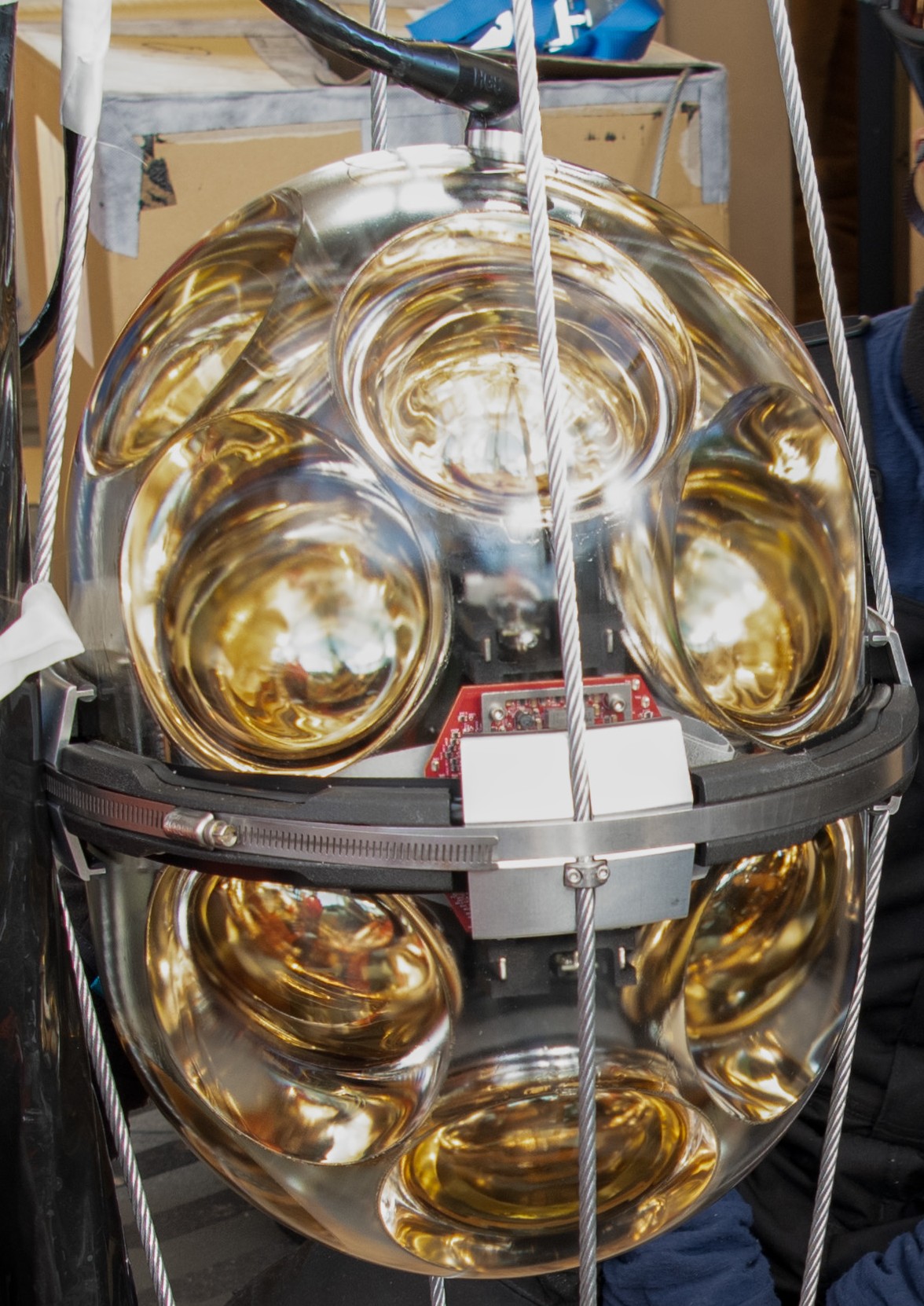}
      \caption{Gen2 DOM}
        \label{fig:lom}
    \end{subfigure}%\hfill
    \begin{subfigure}[b]{0.2116\linewidth}
    \centering
      \includegraphics[width=\scaler\linewidth]{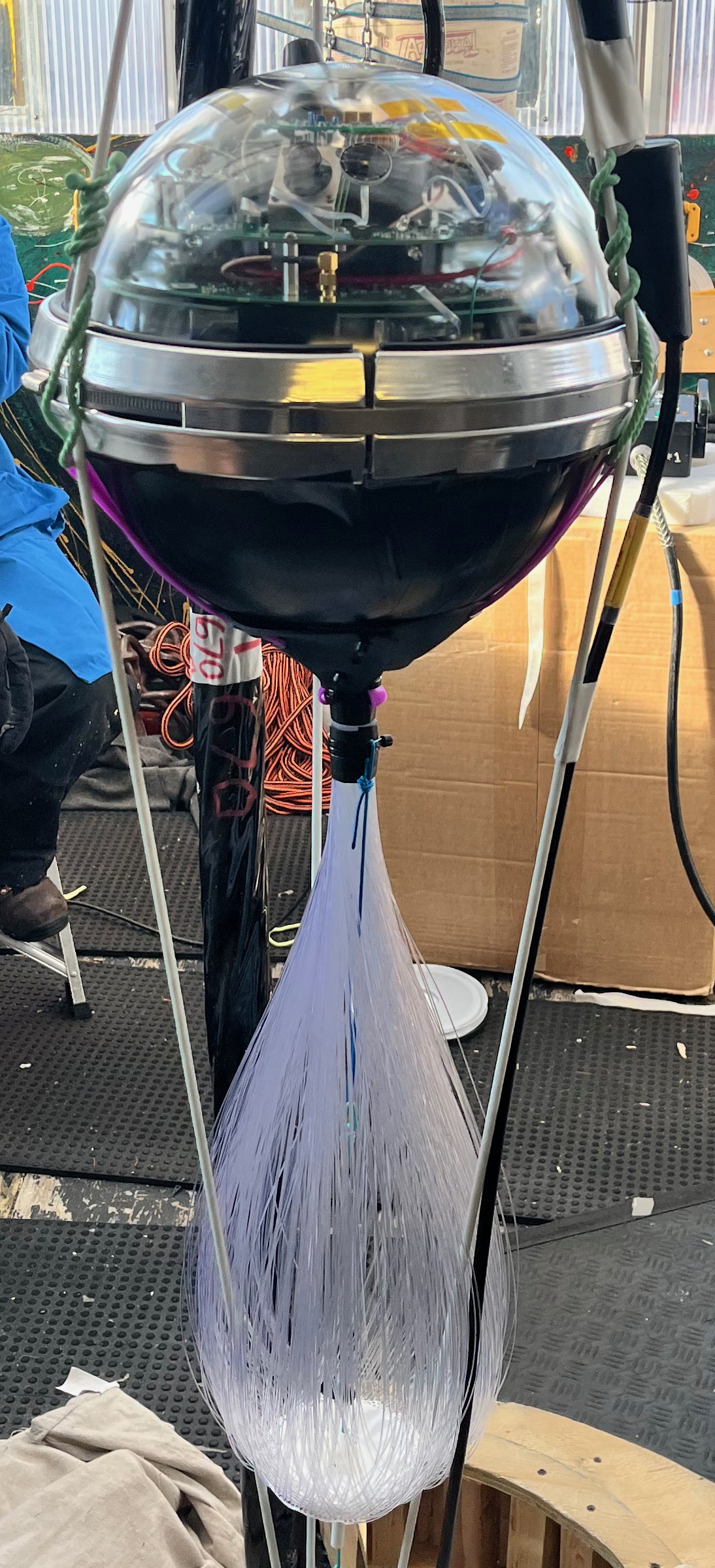}
      \caption{FOM}
        \label{fig:fom}
    \end{subfigure}
  \end{minipage}%\hfill
  %--------- rechter Block: hohes Bild ----------
  \begin{minipage}[t]{0.2\textwidth}
    \vspace{0pt}
    \begin{subfigure}[t]{0.89\linewidth}
    \centering
      \includegraphics[width=\scaler\linewidth]{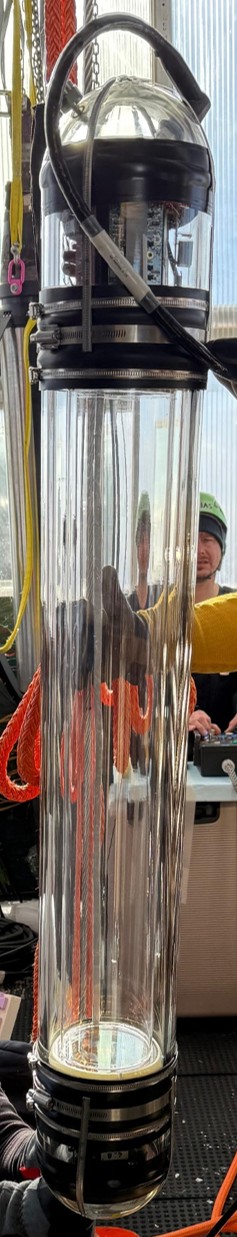}
      \caption{WOM}
        \label{fig:wom}
    \end{subfigure}
  \end{minipage}
  \caption{Photosensor modules integrated in the IceCube Upgrade. The mDOM (a) and D-Egg (b) are the main sensors. The modules shown in (c) to (f) are rarer prototype devices.}
    \label{fig:photo_sensors}
\end{figure}

In total, \num{505} new modules were added to the existing detector. The two main sensor designs are called mDOM \cite{mDOM} and D-Egg \cite{DEgg}. The multi-PMT DOM (mDOM) has \num{24} \SI{3}{inch} PMTs, with angular sensitivity in every direction (see \autoref{fig:mdom}). The Dual optical sensor in Ellipsoid Glass for Gen2 (D-Egg) is equipped with one \SI{8}{inch} PMT facing down and one \SI{8}{inch} PMT facing up (see \autoref{fig:degg}).

Besides the two main sensors, other design ideas are being tested. The old Gen1 DOM design was retro-fitted and new calibration capabilities, such as cameras and acoustic sensors, were added (see \autoref{fig:pdom}). Also, prototype versions of modules equipped with \num{16} or \num{18} PMTs \cite{lom} for the next expansion of IceCube, called IceCube-Gen2 \cite{Gen2paper}, are being tested (see \autoref{fig:lom}). In addition, concepts for Fiber Optical Modules (FOM) (see \autoref{fig:fom}) and Wavelength shifting Optical Modules (WOM) \cite{wom} (see \autoref{fig:wom}) are being tested. There are about \num{10} modules per special device type distributed over all strings and depths.

\subsection{Deployed calibration devices}\label{calibration_devices}

\begin{figure}[htbp]
  \centering
  %--------- linker Block: zwei Zeilen ----------
  \begin{minipage}[t]{0.79\textwidth}
    \vspace{0pt}% Oberkanten der beiden Minipages ausrichten
    % Zeile 1: zwei Querformat-Bilder
    \begin{subfigure}[b]{0.48\linewidth}
    \centering
      \includegraphics[width=0.95\linewidth]{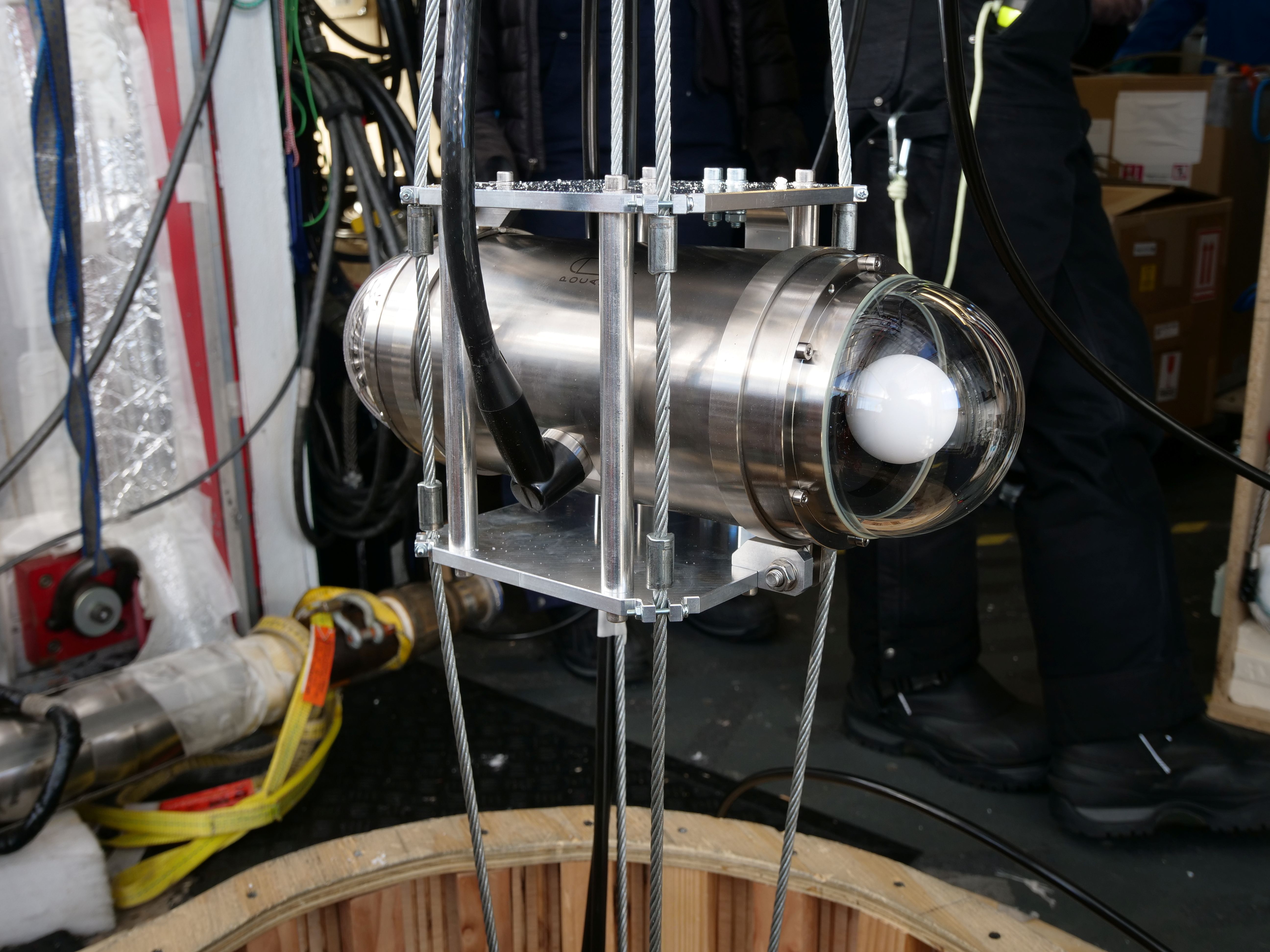}
      \caption{POCAM}
        \label{fig:pocam}
    \end{subfigure}\hfill
    \begin{subfigure}[b]{0.48\linewidth}
    \centering
      \includegraphics[width=0.56\linewidth]{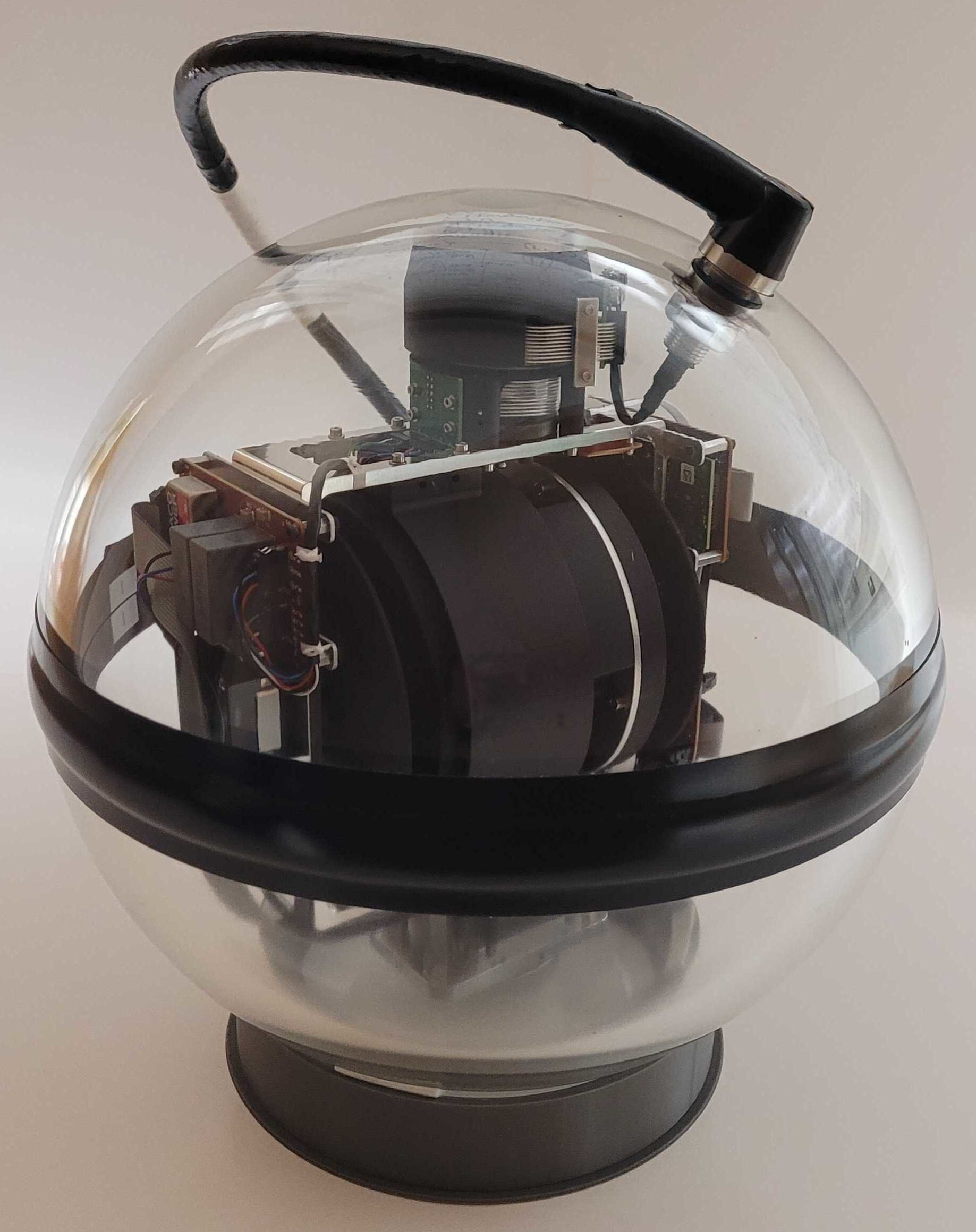}
      \caption{Pencil Beam}
        \label{fig:pencil_beam}
    \end{subfigure}

    \vspace{1ex}

    % Zeile 2: drei Querformat-Bilder
    \begin{subfigure}[b]{0.31\linewidth}
        \centering
      \includegraphics[width=0.93\linewidth]{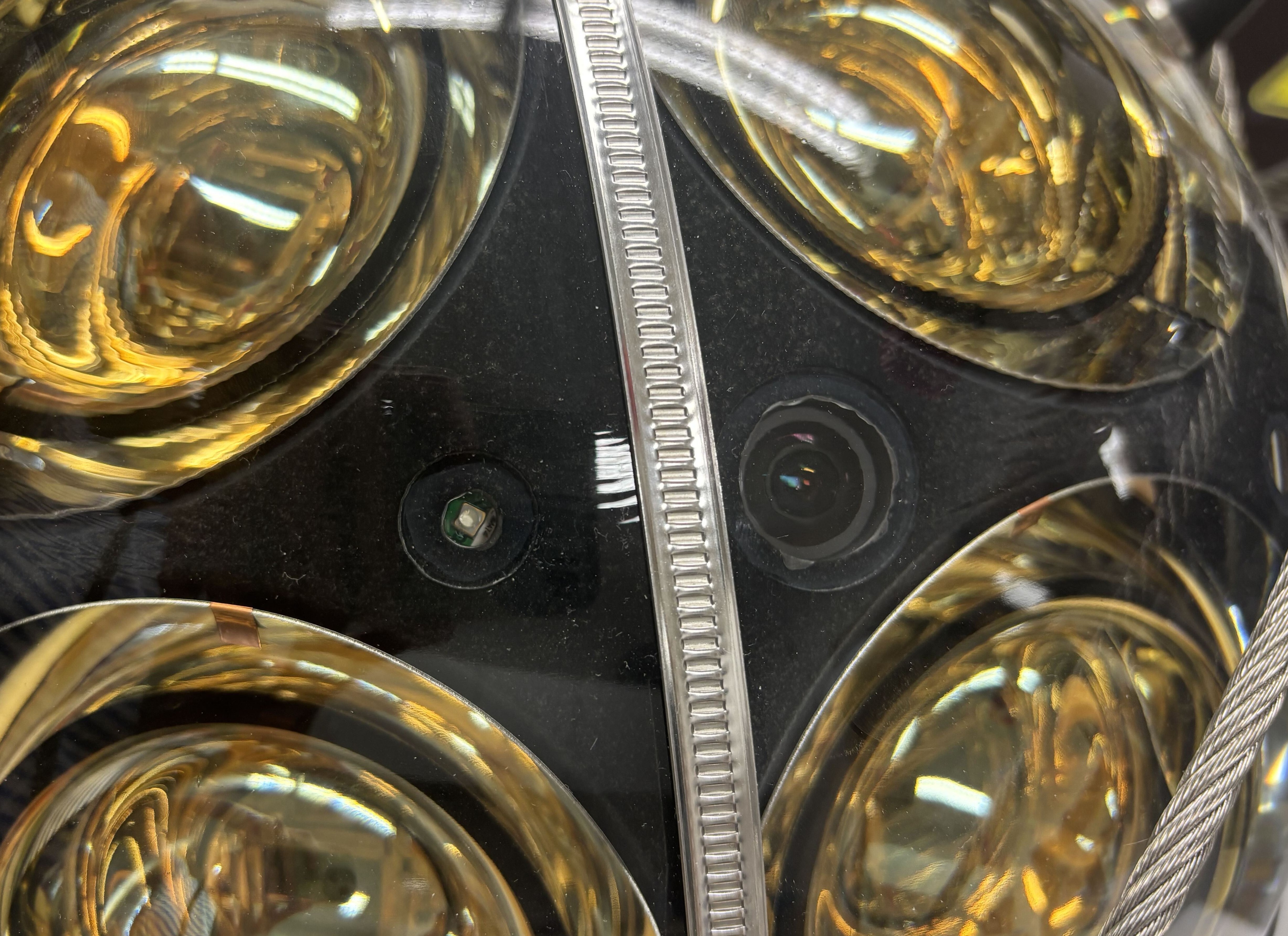}
      \caption{Fixed focus cameras}
        \label{fig:ffc}
    \end{subfigure}\hfill
    \begin{subfigure}[b]{0.31\linewidth}
    \centering
      \includegraphics[width=\linewidth]{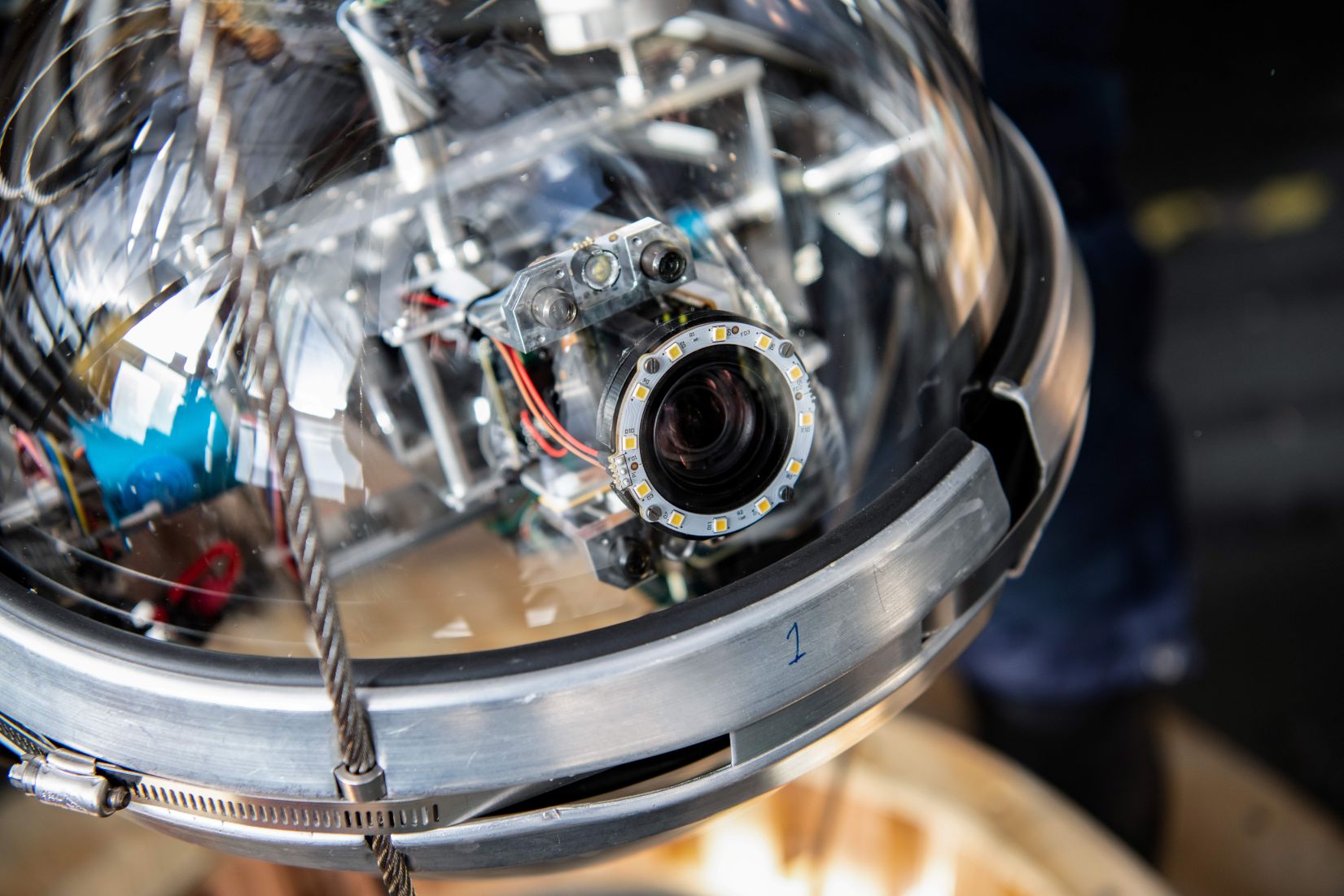}
      \caption{Sweden Camera}
        \label{fig:swecam}
    \end{subfigure}\hfill
    \begin{subfigure}[b]{0.31\linewidth}
    \centering
      \includegraphics[width=\linewidth]{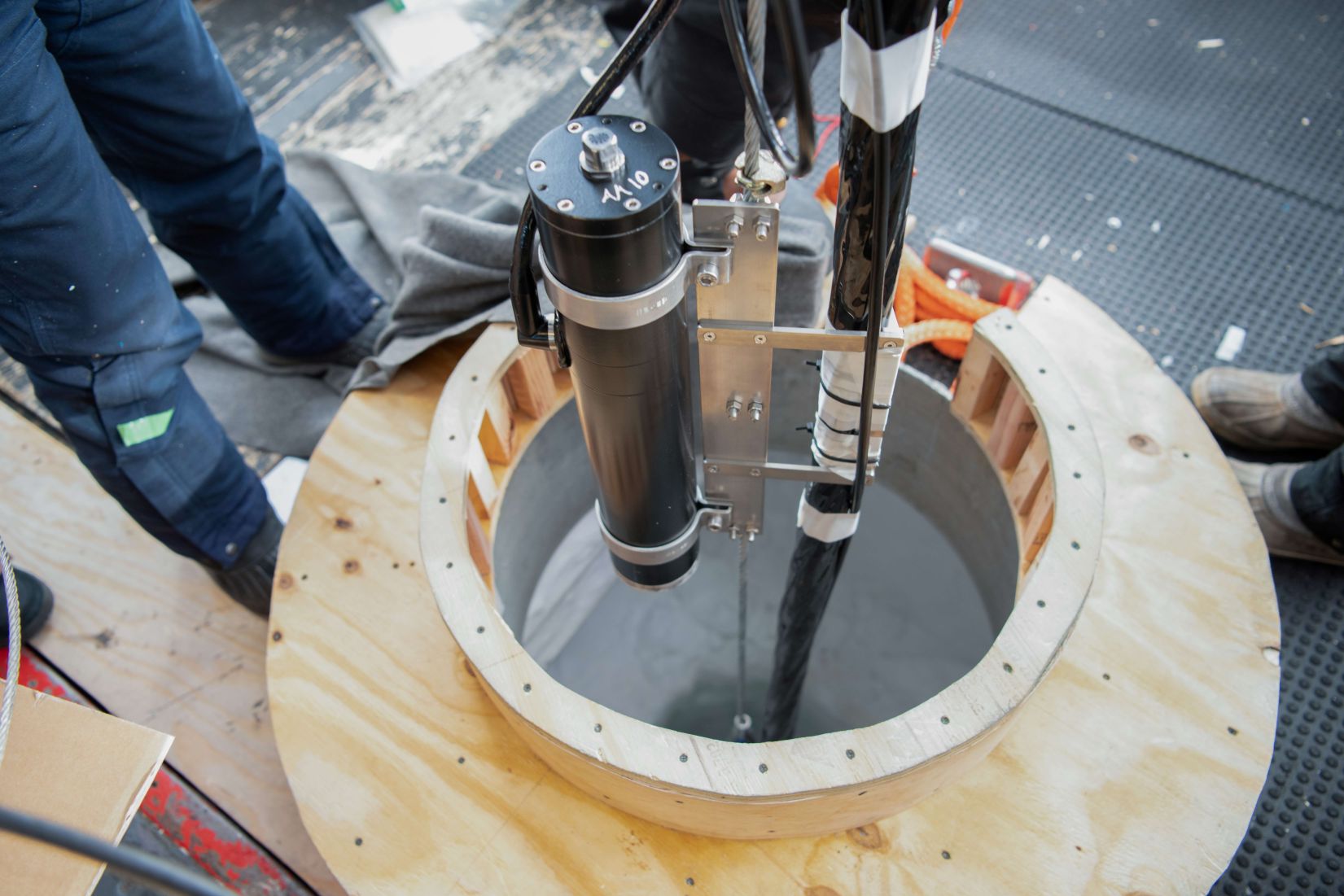}
      \caption{Acoustic Module}
        \label{fig:am}
    \end{subfigure}
  \end{minipage}\hfill
  %--------- rechter Block: hohes Bild ----------
  \begin{minipage}[t]{0.2\textwidth}
    \vspace{0pt}
    \begin{subfigure}[t]{\linewidth}
    \centering
      \includegraphics[width=0.34\linewidth]{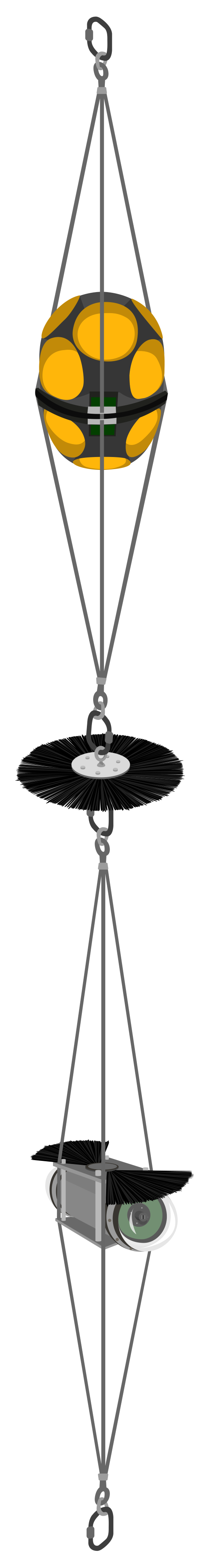}
      \caption{LOMlogger}
        \label{fig:lomlogger}
    \end{subfigure}
  \end{minipage}

  \caption{Calibration modules integrated in the IceCube Upgrade. }
    \label{fig:calibration_modules}
\end{figure}

With the IceCube Upgrade, an improved understanding of the optical properties of the glacial ice is anticipated. Effects such as the ice optical anisotropy \cite{anisotropy}, and of the ice forming the refrozen drill holes, called hole ice, are studied in detail. To be able to gain new knowledge about these effects, several calibration devices were proposed. Their capabilities range from isotropic and directional light sources over different camera systems to acoustic emitters and receivers. The purposes of each device shown in \autoref{fig:calibration_modules} will be discussed in greater detail in the following paragraphs. %in \autoref{calibration_capabilities}.

%The hole ice originates from the \SI{2450}{m} deep holes, which were melted into the glacier to be able to install the instrumentation. Water was kept inside the boreholes that refroze after the strings were lowered. To understand the differences between refrozen ice and the rest of the glacier and to model the hole ice, calibration is needed. 

%\section{Detector Perspectives}\label{perspectives}

%The goal to design new photosensors is archived. The sensors capabilities will be tested with the aim to have better sensitivities for neutrino oscillations. To enhance the neutrino oscillation capabilities of the detector even further, the ice calibration needs to be refined. The refinement of the calibration is currently in progress and the plans will be explored in the following chapter, \autoref{calibration_capabilities}. The long term goal of neutrino oscillations will be further discussed in \autoref{oscillation_capabilities}.

%\subsection{Calibration capabilities}\label{calibration_capabilities}
%(ZITATE EINFÜGEN!)

Every mDOM and D-Egg contains \num{3} fixed-focus cameras \cite{ffc} (see \autoref{fig:ffc}), which can record images of the hole ice and neighboring modules on the string. These cameras can, among other things, deliver insight about the development of the high scattering column, called bubble column, within the refreezing hole.

A fully steerable camera, called Sweden Camera (see \autoref{fig:swecam}), is able to record videos and images of the surrounding hole ice and how it refreezes. So far, it has been able to capture ice crystal formation processes and the bubble column formation.

The Pencil Beam is a steerable light source (see \autoref{fig:pencil_beam}). It emits a collimated light beam chosen from one of 16 LEDs in 8 different wavelengths or one laser. The device has a full movement range in azimuthal and polar direction, which is needed to further investigate the optical ice anisotropy \cite{pencil_beam}. 

The isotropic light source, called Precision Optical CAlibration Module (POCAM) \cite{POCAM} (see \autoref{fig:pocam}), will be able to measure the DOM efficiency as well as hole ice properties. In addition, it was used during its deployment, while lowering the instrumentation, to measure the angular acceptance of Gen1 DOMs at various depths. The expectation of this measurement is described in \cite{POCAM_angular_acceptance}.

Another depth-dependent measurement was the study of scattering properties of the different ice layers within the glacier. The LOMlogger \cite{lomlogger} (see \autoref{fig:lomlogger}) used a modified POCAM to shine light into the ice. The backscattered light was detected by one of the Gen2-prototype detector modules. This new measurement device delivered results comparable to its predecessor instrument and can be used for IceCube-Gen2.

The acoustic module \cite{acoustic} is another calibration effort leading towards IceCube-Gen2. The modules deployed in the IceCube Upgrade, consisting of acoustic emitters and receivers, will test acoustic ice properties, such as speed of sound and attenuation. These measurements will inform the feasibility of acoustic geometry calibration at the \SI{250}{m} string spacing planned for IceCube-Gen2.

\subsection{Physics capabilities}\label{oscillation_capabilities}

The new sensor modules and the narrow spacing of the new strings improve IceCube's sensitivity to measure neutrino oscillation. First sensitivity studies presented in \cite{oscillations} suggest that, due to the additional strings, the mass hierarchy can be determined with a \num{3}$\sigma$ significance within \num{5} years in case of normal ordering. Based on preliminary indications, the impact of having two fewer strings compared to the \num{7}-string detector layout assumed in this study is expected to be modest.

Besides oscillations, the new strings will allow an improved sensitivity for low-energy astrophysics \cite{astrophysical_sensitivity_projections_upgrade}, such as \si{GeV} neutrino transient studies. Also, further beyond-the-standard-model searches will be possible, for example, low-mass dark matter searches \cite{abbasi2026sensitivityprojectionslowmassdark}.

Additionally, the new calibration devices, discussed above, will improve the model of the light propagation in ice. This is expected to lead to a better neutrino event reconstruction, which can also be applied retroactively to the more than \num{15} years of archival data. %The impact of previous ice model improvements were studied in \cite{improved_shower_modeling}.

\section{Conclusion}

%The status of the detector is that \num{5} new strings are about to take data with the rest of IceCube in fall 2026. 
The recently deployed IceCube Upgrade is presently in its commissioning phase before its integration into the IceCube data acquisition. New calibration campaigns are ongoing. The measurements studying the refreeze of the hole ice are currently evaluated. Ice property measurements, such as scattering, hole ice, and anisotropy, are currently performed. New geometry and timing calibrations are also an ongoing effort. In the future, all new calibration results will be applicable to new measurements as well as to archival data and will lead to more precise reconstruction and pointing accuracy.

%For one-column wide figures use syntax of figure~\ref{fig-1}
%\begin{figure}[h]
%\centering
%\includegraphics[width=5cm,clip]{fig-1-sample}
%\label{fig-1}       % Give a unique label
%\end{figure}

%For two-column wide figures use syntax of figure~\ref{fig-2}
%\begin{figure*}
%\centering
%\vspace*{1cm}       % Give the correct figure height in cm
%\includegraphics[width=8cm,clip]{fig-2-sample}
%\label{fig-2}       % Give a unique label
%\end{figure*}

%For figure with sidecaption legend use syntax of figure~\ref{fig-3}
%\begin{figure}
%\centering
%\sidecaption
%\includegraphics[width=5cm,clip]{fig-3-sample}
%\label{fig-3}       % Give a unique label
%\end{figure}

%

% BibTeX or Biber users please use (the style is already called in the class, ensure that the "woc.bst" style is in your local directory)
\bibliography{bibliography} % Replace "your_bib_file" with the actual name of your .bib file

@article{IceCube_instrumentation_and_online_system,
	doi = {10.1088/1748-0221/12/03/p03012},
    collaboration = "IceCube Collaboration",
	year = {2017},
	month = {03},
	publisher = {{IOP} Publishing},
	volume = {12}, 
	number = {03},
	pages = {P03012--P03012},
	author = {M.G. Aartsen and others},
    IGNOREauthor = {M.G. Aartsen and M. Ackermann and J. Adams and J.A. Aguilar and M. Ahlers and M. Ahrens and D. Altmann and K. Andeen and T. Anderson and I. Ansseau and G. Anton and M. Archinger and C. Argüelles and R. Auer and J. Auffenberg and S. Axani and J. Baccus and X. Bai and S. Barnet and S.W. Barwick and V. Baum and R. Bay and K. Beattie and J.J. Beatty and J. Becker Tjus and K.-H. Becker and T. Bendfelt and S. BenZvi and D. Berley and E. Bernardini and A. Bernhard and D.Z. Besson and G. Binder and D. Bindig and M. Bissok and E. Blaufuss and S. Blot and D. Boersma and C. Bohm and M. Börner and F. Bos and D. Bose and S. Böser and O. Botner and A. Bouchta and J. Braun and L. Brayeur and H.-P. Bretz and S. Bron and A. Burgman and C. Burreson and T. Carver and M. Casier and E. Cheung and D. Chirkin and A. Christov and K. Clark and L. Classen and S. Coenders and G.H. Collin and J.M. Conrad and D.F. Cowen and R. Cross and C. Day and M. Day and J.P.A.M. de Andr{\'{e}} and C. De Clercq and E. del Pino Rosendo and H. Dembinski and S. De Ridder and F. Descamps and P. Desiati and K.D. de Vries and G. de Wasseige and M. de With and T. DeYoung and J.C. D{\'{\i}}az-V{\'{e}}lez and V. di Lorenzo and H. Dujmovic and J.P. Dumm and M. Dunkman and B. Eberhardt and W.R. Edwards and T. Ehrhardt and B. Eichmann and P. Eller and S. Euler and P.A. Evenson and S. Fahey and A.R. Fazely and J. Feintzeig and J. Felde and K. Filimonov and C. Finley and S. Flis and C.-C. Fösig and A. Franckowiak and M. Fr{\`{e}}re and E. Friedman and T. Fuchs and T.K. Gaisser and J. Gallagher and L. Gerhardt and K. Ghorbani and W. Giang and L. Gladstone and T. Glauch and D. Glowacki and T. Glüsenkamp and A. Goldschmidt and J.G. Gonzalez and D. Grant and Z. Griffith and L. Gustafsson and C. Haack and A. Hallgren and F. Halzen and E. Hansen and T. Hansmann and K. Hanson and J. Haugen and D. Hebecker and D. Heereman and K. Helbing and R. Hellauer and R. Heller and S. Hickford and J. Hignight and G.C. Hill and K.D. Hoffman and R. Hoffmann and K. Hoshina and F. Huang and M. Huber and P.O. Hulth and K. Hultqvist and S. In and M. Inaba and A. Ishihara and E. Jacobi and J. Jacobsen and G.S. Japaridze and M. Jeong and K. Jero and A. Jones and B.J.P. Jones and J. Joseph and W. Kang and A. Kappes and T. Karg and A. Karle and U. Katz and M. Kauer and A. Keivani and J.L. Kelley and J. Kemp and A. Kheirandish and J. Kim and M. Kim and T. Kintscher and J. Kiryluk and N. Kitamura and T. Kittler and S.R. Klein and S. Kleinfelder and M. Kleist and G. Kohnen and R. Koirala and H. Kolanoski and R. Konietz and L. Köpke and C. Kopper and S. Kopper and D.J. Koskinen and M. Kowalski and M. Krasberg and K. Krings and M. Kroll and G. Krückl and C. Krüger and J. Kunnen and S. Kunwar and N. Kurahashi and T. Kuwabara and M. Labare and K. Laihem and H. Landsman and J.L. Lanfranchi and M.J. Larson and F. Lauber and A. Laundrie and D. Lennarz and H. Leich and M. Lesiak-Bzdak and M. Leuermann and L. Lu and J. Ludwig and J. Lünemann and C. Mackenzie and J. Madsen and G. Maggi and K.B.M. Mahn and S. Mancina and M. Mandelartz and R. Maruyama and K. Mase and H. Matis and R. Maunu and F. McNally and C.P. McParland and P. Meade and K. Meagher and M. Medici and M. Meier and A. Meli and T. Menne and G. Merino and T. Meures and S. Miarecki and R.H. Minor and T. Montaruli and M. Moulai and T. Murray and R. Nahnhauer and U. Naumann and G. Neer and M. Newcomb and H. Niederhausen and S.C. Nowicki and D.R. Nygren and A. Obertacke Pollmann and A. Olivas and A. O{\textquotesingle}Murchadha and T. Palczewski and H. Pandya and D.V. Pankova and S. Patton and P. Peiffer and Ö. Penek and J.A. Pepper and C. P{\'{e}}rez de los Heros and C. Pettersen and D. Pieloth and E. Pinat and P.B. Price and G.T. Przybylski and M. Quinnan and C. Raab and L. Rädel and M. Rameez and K. Rawlins and R. Reimann and B. Relethford and M. Relich and E. Resconi and W. Rhode and M. Richman and B. Riedel and S. Robertson and M. Rongen and C. Roucelle and C. Rott and T. Ruhe and D. Ryckbosch and D. Rysewyk and L. Sabbatini and S.E. Sanchez Herrera and A. Sandrock and J. Sandroos and P. Sandstrom and S. Sarkar and K. Satalecka and P. Schlunder and T. Schmidt and S. Schoenen and S. Schöneberg and A. Schukraft and L. Schumacher and D. Seckel and S. Seunarine and M. Solarz and D. Soldin and M. Song and G.M. Spiczak and C. Spiering and T. Stanev and A. Stasik and J. Stettner and A. Steuer and T. Stezelberger and R.G. Stokstad and A. Stö{\ss}l and R. Ström and N.L. Strotjohann and K.-H. Sulanke and G.W. Sullivan and M. Sutherland and H. Taavola and I. Taboada and J. Tatar and F. Tenholt and S. Ter-Antonyan and A. Terliuk and G. Te{\v{s}}i{\'{c}} and L. Thollander and S. Tilav and P.A. Toale and M.N. Tobin and S. Toscano and D. Tosi and M. Tselengidou and A. Turcati and E. Unger and M. Usner and J. Vandenbroucke and N. van Eijndhoven and S. Vanheule and M. van Rossem and J. van Santen and M. Vehring and M. Voge and E. Vogel and M. Vraeghe and D. Wahl and C. Walck and A. Wallace and M. Wallraff and N. Wandkowsky and Ch. Weaver and M.J. Weiss and C. Wendt and S. Westerhoff and D. Wharton and B.J. Whelan and S. Wickmann and K. Wiebe and C.H. Wiebusch and L. Wille and D.R. Williams and L. Wills and P. Wisniewski and M. Wolf and T.R. Wood and E. Woolsey and K. Woschnagg and D.L. Xu and X.W. Xu and Y. Xu and J.P. Yanez and G. Yodh and S. Yoshida and M. Zoll},
	title = "{The {IceCube} Neutrino Observatory: instrumentation and online systems}", 
	journal = {Journal of Instrumentation}
}

@article{IceCube:2011ucd,
    author = "Abbasi, R. and others",
    collaboration = "IceCube Collaboration",
    title = "{The Design and Performance of IceCube DeepCore}",
    eprint = "1109.6096",
    archivePrefix = "arXiv",
    primaryClass = "astro-ph.IM",
    doi = "10.1016/j.astropartphys.2012.01.004",
    journal = "Astropart. Phys.",
    volume = "35",
    pages = "615--624",
    year = "2012"
}

@article{
galactic_plane,
author = {R. Abbasi and others},
IGNOREauthor = {IceCube Collaboration*† and R. Abbasi and M. Ackermann  and J. Adams  and J. A. Aguilar  and M. Ahlers  and M. Ahrens  and J. M. Alameddine  and A. A. Alves  and N. M. Amin  and K. Andeen  and T. Anderson  and G. Anton  and C. Argüelles  and Y. Ashida  and S. Athanasiadou  and S. Axani  and X. Bai  and A. Balagopal V.  and S. W. Barwick  and V. Basu  and S. Baur  and R. Bay  and J. J. Beatty  and K.-H. Becker  and J. Becker Tjus  and J. Beise  and C. Bellenghi  and S. Benda  and S. BenZvi  and D. Berley  and E. Bernardini  and D. Z. Besson  and G. Binder  and D. Bindig  and E. Blaufuss  and S. Blot  and M. Boddenberg  and F. Bontempo  and J. Y. Book  and J. Borowka  and S. Böser  and O. Botner  and J. Böttcher  and E. Bourbeau  and F. Bradascio  and J. Braun  and B. Brinson  and S. Bron  and J. Brostean-Kaiser  and R. T. Burley  and R. S. Busse  and M. A. Campana  and E. G. Carnie-Bronca  and C. Chen  and Z. Chen  and D. Chirkin  and K. Choi  and B. A. Clark  and K. Clark  and L. Classen  and A. Coleman  and G. H. Collin  and A. Connolly  and J. M. Conrad  and P. Coppin  and P. Correa  and D. F. Cowen  and R. Cross  and C. Dappen  and P. Dave  and C. De Clercq  and J. J. DeLaunay  and D. Delgado López  and H. Dembinski  and K. Deoskar  and A. Desai  and P. Desiati  and K. D. de Vries  and G. de Wasseige  and T. DeYoung  and A. Diaz  and J. C. Díaz-Vélez  and M. Dittmer  and H. Dujmovic  and M. Dunkman  and M. A. DuVernois  and T. Ehrhardt  and P. Eller  and R. Engel  and H. Erpenbeck  and J. Evans  and P. A. Evenson  and K. L. Fan  and A. R. Fazely  and A. Fedynitch  and N. Feigl  and S. Fiedlschuster  and A. T. Fienberg  and C. Finley  and L. Fischer  and D. Fox  and A. Franckowiak  and E. Friedman  and A. Fritz  and P. Fürst  and T. K. Gaisser  and J. Gallagher  and E. Ganster  and A. Garcia  and S. Garrappa  and L. Gerhardt  and A. Ghadimi  and C. Glaser  and T. Glauch  and T. Glüsenkamp  and N. Goehlke  and A. Goldschmidt  and J. G. Gonzalez  and S. Goswami  and D. Grant  and T. Grégoire  and S. Griswold  and C. Günther  and P. Gutjahr  and C. Haack  and A. Hallgren  and R. Halliday  and L. Halve  and F. Halzen  and M. Ha Minh  and K. Hanson  and J. Hardin  and A. A. Harnisch  and A. Haungs  and K. Helbing  and F. Henningsen  and E. C. Hettinger  and S. Hickford  and J. Hignight  and C. Hill  and G. C. Hill  and K. D. Hoffman  and K. Hoshina  and W. Hou  and F. Huang  and M. Huber  and T. Huber  and K. Hultqvist  and M. Hünnefeld  and R. Hussain  and K. Hymon  and S. In  and N. Iovine  and A. Ishihara  and M. Jansson  and G. S. Japaridze  and M. Jeong  and M. Jin  and B. J. P. Jones  and D. Kang  and W. Kang  and X. Kang  and A. Kappes  and D. Kappesser  and L. Kardum  and T. Karg  and M. Karl  and A. Karle  and U. Katz  and M. Kauer  and M. Kellermann  and J. L. Kelley  and A. Kheirandish  and K. Kin  and J. Kiryluk  and S. R. Klein  and A. Kochocki  and R. Koirala  and H. Kolanoski  and T. Kontrimas  and L. Köpke  and C. Kopper  and S. Kopper  and D. J. Koskinen  and P. Koundal  and M. Kovacevich  and M. Kowalski  and T. Kozynets  and E. Krupczak  and E. Kun  and N. Kurahashi  and N. Lad  and C. Lagunas Gualda  and J. L. Lanfranchi  and M. J. Larson  and F. Lauber  and J. P. Lazar  and J. W. Lee  and K. Leonard  and A. Leszczyńska  and Y. Li  and M. Lincetto  and Q. R. Liu  and M. Liubarska  and E. Lohfink  and C. J. Lozano Mariscal  and L. Lu  and F. Lucarelli  and A. Ludwig  and W. Luszczak  and Y. Lyu  and W. Y. Ma  and J. Madsen  and K. B. M. Mahn  and Y. Makino  and S. Mancina  and I. C. Mariş  and I. Martinez-Soler  and R. Maruyama  and S. McHale  and T. McElroy  and F. McNally  and J. V. Mead  and K. Meagher  and S. Mechbal  and A. Medina  and M. Meier  and S. Meighen-Berger  and Y. Merckx  and J. Micallef  and D. Mockler  and T. Montaruli  and R. W. Moore  and K. Morik  and R. Morse  and M. Moulai  and T. Mukherjee  and R. Naab  and R. Nagai  and R. Nahnhauer  and U. Naumann  and J. Necker  and L. V. Nguyen  and H. Niederhausen  and M. U. Nisa  and S. C. Nowicki  and D. Nygren  and A. Obertacke Pollmann  and M. Oehler  and B. Oeyen  and A. Olivas  and E. O'Sullivan  and H. Pandya  and D. V. Pankova  and N. Park  and G. K. Parker  and E. N. Paudel  and L. Paul  and C. Pérez de los Heros  and L. Peters  and J. Peterson  and S. Philippen  and S. Pieper  and A. Pizzuto  and M. Plum  and Y. Popovych  and A. Porcelli  and M. Prado Rodriguez  and B. Pries  and G. T. Przybylski  and C. Raab  and J. Rack-Helleis  and A. Raissi  and M. Rameez  and K. Rawlins  and I. C. Rea  and Z. Rechav  and A. Rehman  and P. Reichherzer  and R. Reimann  and G. Renzi  and E. Resconi  and S. Reusch  and W. Rhode  and M. Richman  and B. Riedel  and E. J. Roberts  and S. Robertson  and G. Roellinghoff  and M. Rongen  and C. Rott  and T. Ruhe  and D. Ryckbosch  and D. Rysewyk Cantu  and I. Safa  and J. Saffer  and D. Salazar-Gallegos  and P. Sampathkumar  and S. E. Sanchez Herrera  and A. Sandrock  and M. Santander  and S. Sarkar  and S. Sarkar  and K. Satalecka  and M. Schaufel  and H. Schieler  and S. Schindler  and T. Schmidt  and A. Schneider  and J. Schneider  and F. G. Schröder  and L. Schumacher  and G. Schwefer  and S. Sclafani  and D. Seckel  and S. Seunarine  and A. Sharma  and S. Shefali  and N. Shimizu  and M. Silva  and B. Skrzypek  and B. Smithers  and R. Snihur  and J. Soedingrekso  and A. Sogaard  and D. Soldin  and C. Spannfellner  and G. M. Spiczak  and C. Spiering  and M. Stamatikos  and T. Stanev  and R. Stein  and J. Stettner  and T. Stezelberger  and B. Stokstad  and T. Stürwald  and T. Stuttard  and G. W. Sullivan  and I. Taboada  and S. Ter-Antonyan  and J. Thwaites  and S. Tilav  and F. Tischbein  and K. Tollefson  and C. Tönnis  and S. Toscano  and D. Tosi  and A. Trettin  and M. Tselengidou  and C. F. Tung  and A. Turcati  and R. Turcotte  and C. F. Turley  and J. P. Twagirayezu  and B. Ty  and M. A. Unland Elorrieta  and N. Valtonen-Mattila  and J. Vandenbroucke  and N. van Eijndhoven  and D. Vannerom  and J. van Santen  and J. Veitch-Michaelis  and S. Verpoest  and C. Walck  and W. Wang  and T. B. Watson  and C. Weaver  and P. Weigel  and A. Weindl  and M. J. Weiss  and J. Weldert  and C. Wendt  and J. Werthebach  and M. Weyrauch  and N. Whitehorn  and C. H. Wiebusch  and N. Willey  and D. R. Williams  and M. Wolf  and G. Wrede  and J. Wulff  and X. W. Xu  and J. P. Yanez  and E. Yildizci  and S. Yoshida  and S. Yu  and T. Yuan  and Z. Zhang  and P. Zhelnin },
collaboration = {IceCube Collaboration},
title = "{Observation of high-energy neutrinos from the Galactic plane}",
journal = {Science},
volume = {380},
number = {6652},
pages = {1338-1343},
year = {2023},
doi = {10.1126/science.adc9818},
IGNOREURL = {https://www.science.org/doi/abs/10.1126/science.adc9818},
IGNOREeprint = {https://www.science.org/doi/pdf/10.1126/science.adc9818}}

@article{
ngc1068,
author = {R. Abbasi and others},
IGNOREauthor = {IceCube Collaboration*† and R. Abbasi and M. Ackermann  and J. Adams  and J. A. Aguilar  and M. Ahlers  and M. Ahrens  and J. M. Alameddine  and C. Alispach  and A. A. Alves  and N. M. Amin  and K. Andeen  and T. Anderson  and G. Anton  and C. Argüelles  and Y. Ashida  and S. Axani  and X. Bai  and A. Balagopal V.  and A. Barbano  and S. W. Barwick  and B. Bastian  and V. Basu  and S. Baur  and R. Bay  and J. J. Beatty  and K.-H. Becker  and J. Becker Tjus  and C. Bellenghi  and S. BenZvi  and D. Berley  and E. Bernardini  and D. Z. Besson  and G. Binder  and D. Bindig  and E. Blaufuss  and S. Blot  and M. Boddenberg  and F. Bontempo  and J. Borowka  and S. Böser  and O. Botner  and J. Böttcher  and E. Bourbeau  and F. Bradascio  and J. Braun  and B. Brinson  and S. Bron  and J. Brostean-Kaiser  and S. Browne  and A. Burgman  and R. T. Burley  and R. S. Busse  and M. A. Campana  and E. G. Carnie-Bronca  and C. Chen  and Z. Chen  and D. Chirkin  and K. Choi  and B. A. Clark  and K. Clark  and L. Classen  and A. Coleman  and G. H. Collin  and J. M. Conrad  and P. Coppin  and P. Correa  and D. F. Cowen  and R. Cross  and C. Dappen  and P. Dave  and C. De Clercq  and J. J. DeLaunay  and D. Delgado López  and H. Dembinski  and K. Deoskar  and A. Desai  and P. Desiati  and K. D. de Vries  and G. de Wasseige  and M. de With  and T. DeYoung  and A. Diaz  and J. C. Díaz-Vélez  and M. Dittmer  and H. Dujmovic  and M. Dunkman  and M. A. DuVernois  and E. Dvorak  and T. Ehrhardt  and P. Eller  and R. Engel  and H. Erpenbeck  and J. Evans  and P. A. Evenson  and K. L. Fan  and A. R. Fazely  and A. Fedynitch  and N. Feigl  and S. Fiedlschuster  and A. T. Fienberg  and K. Filimonov  and C. Finley  and L. Fischer  and D. Fox  and A. Franckowiak  and E. Friedman  and A. Fritz  and P. Fürst  and T. K. Gaisser  and J. Gallagher  and E. Ganster  and A. Garcia  and S. Garrappa  and L. Gerhardt  and A. Ghadimi  and C. Glaser  and T. Glauch  and T. Glüsenkamp  and A. Goldschmidt  and J. G. Gonzalez  and S. Goswami  and D. Grant  and T. Grégoire  and S. Griswold  and C. Günther  and P. Gutjahr  and C. Haack  and A. Hallgren  and R. Halliday  and L. Halve  and F. Halzen  and M. Ha Minh  and K. Hanson  and J. Hardin  and A. A. Harnisch  and A. Haungs  and D. Hebecker  and K. Helbing  and F. Henningsen  and E. C. Hettinger  and S. Hickford  and J. Hignight  and C. Hill  and G. C. Hill  and K. D. Hoffman  and R. Hoffmann  and B. Hokanson-Fasig  and K. Hoshina  and F. Huang  and M. Huber  and T. Huber  and K. Hultqvist  and M. Hünnefeld  and R. Hussain  and K. Hymon  and S. In  and N. Iovine  and A. Ishihara  and M. Jansson  and G. S. Japaridze  and M. Jeong  and M. Jin  and B. J. P. Jones  and D. Kang  and W. Kang  and X. Kang  and A. Kappes  and D. Kappesser  and L. Kardum  and T. Karg  and M. Karl  and A. Karle  and U. Katz  and M. Kauer  and M. Kellermann  and J. L. Kelley  and A. Kheirandish  and K. Kin  and T. Kintscher  and J. Kiryluk  and S. R. Klein  and R. Koirala  and H. Kolanoski  and T. Kontrimas  and L. Köpke  and C. Kopper  and S. Kopper  and D. J. Koskinen  and P. Koundal  and M. Kovacevich  and M. Kowalski  and T. Kozynets  and E. Kun  and N. Kurahashi  and N. Lad  and C. Lagunas Gualda  and J. L. Lanfranchi  and M. J. Larson  and F. Lauber  and J. P. Lazar  and J. W. Lee  and K. Leonard  and A. Leszczyńska  and Y. Li  and M. Lincetto  and Q. R. Liu  and M. Liubarska  and E. Lohfink  and C. J. Lozano Mariscal  and L. Lu  and F. Lucarelli  and A. Ludwig  and W. Luszczak  and Y. Lyu  and W. Y. Ma  and J. Madsen  and K. B. M. Mahn  and Y. Makino  and S. Mancina  and I. C. Mariş  and I. Martinez-Soler  and R. Maruyama  and K. Mase  and T. McElroy  and F. McNally  and J. V. Mead  and K. Meagher  and S. Mechbal  and A. Medina  and M. Meier  and S. Meighen-Berger  and J. Micallef  and D. Mockler  and T. Montaruli  and R. W. Moore  and R. Morse  and M. Moulai  and R. Naab  and R. Nagai  and R. Nahnhauer  and U. Naumann  and J. Necker  and L. V. Nguyen  and H. Niederhausen  and M. U. Nisa  and S. C. Nowicki  and D. Nygren  and A. Obertacke Pollmann  and M. Oehler  and B. Oeyen  and A. Olivas  and E. O’Sullivan  and H. Pandya  and D. V. Pankova  and N. Park  and G. K. Parker  and E. N. Paudel  and L. Paul  and C. Pérez de los Heros  and L. Peters  and J. Peterson  and S. Philippen  and S. Pieper  and M. Pittermann  and A. Pizzuto  and M. Plum  and Y. Popovych  and A. Porcelli  and M. Prado Rodriguez  and P. B. Price  and B. Pries  and G. T. Przybylski  and C. Raab  and J. Rack-Helleis  and A. Raissi  and M. Rameez  and K. Rawlins  and I. C. Rea  and A. Rehman  and P. Reichherzer  and R. Reimann  and G. Renzi  and E. Resconi  and S. Reusch  and W. Rhode  and M. Richman  and B. Riedel  and E. J. Roberts  and S. Robertson  and G. Roellinghoff  and M. Rongen  and C. Rott  and T. Ruhe  and D. Ryckbosch  and D. Rysewyk Cantu  and I. Safa  and J. Saffer  and S. E. Sanchez Herrera  and A. Sandrock  and J. Sandroos  and M. Santander  and S. Sarkar  and S. Sarkar  and K. Satalecka  and M. Schaufel  and H. Schieler  and S. Schindler  and T. Schmidt  and A. Schneider  and J. Schneider  and F. G. Schröder  and L. Schumacher  and G. Schwefer  and S. Sclafani  and D. Seckel  and S. Seunarine  and A. Sharma  and S. Shefali  and M. Silva  and B. Skrzypek  and B. Smithers  and R. Snihur  and J. Soedingrekso  and D. Soldin  and C. Spannfellner  and G. M. Spiczak  and C. Spiering  and J. Stachurska  and M. Stamatikos  and T. Stanev  and R. Stein  and J. Stettner  and A. Steuer  and T. Stezelberger  and R. Stokstad  and T. Stürwald  and T. Stuttard  and G. W. Sullivan  and I. Taboada  and S. Ter-Antonyan  and S. Tilav  and F. Tischbein  and K. Tollefson  and C. Tönnis  and S. Toscano  and D. Tosi  and A. Trettin  and M. Tselengidou  and C. F. Tung  and A. Turcati  and R. Turcotte  and C. F. Turley  and J. P. Twagirayezu  and B. Ty  and M. A. Unland Elorrieta  and N. Valtonen-Mattila  and J. Vandenbroucke  and N. van Eijndhoven  and D. Vannerom  and J. van Santen  and S. Verpoest  and C. Walck  and T. B. Watson  and C. Weaver  and P. Weigel  and A. Weindl  and M. J. Weiss  and J. Weldert  and C. Wendt  and J. Werthebach  and M. Weyrauch  and N. Whitehorn  and C. H. Wiebusch  and D. R. Williams  and M. Wolf  and K. Woschnagg  and G. Wrede  and J. Wulff  and X. W. Xu  and J. P. Yanez  and S. Yoshida  and S. Yu  and T. Yuan  and Z. Zhang  and P. Zhelnin },
collaboration = {IceCube Collaboration},
title = "{Evidence for neutrino emission from the nearby active galaxy NGC 1068}",
journal = {Science},
volume = {378},
number = {6619},
pages = {538-543},
year = {2022},
doi = {10.1126/science.abg3395},
IGNOREURL = {https://www.science.org/doi/abs/10.1126/science.abg3395},
IGNOREeprint = {https://www.science.org/doi/pdf/10.1126/science.abg3395}}

@article{
3year_diffuse_flux,
author = {M. G. Aartsen and others},
collaboration = {IceCube Collaboration},
title = "{Evidence for High-Energy Extraterrestrial Neutrinos at the IceCube Detector}",
journal = {Science},
volume = {342},
number = {6161},
pages = {1242856},
year = {2013},
doi = {10.1126/science.1242856},
IGNOREURL = {https://www.science.org/doi/abs/10.1126/science.1242856},
IGNOREeprint = {https://www.science.org/doi/pdf/10.1126/science.1242856}}

%\begin{thebibliography}{}
%Journal Author, Article title. Journal \textbf{Volume}, page numbers (year). \url{https://doi.org/Article-DOI-number}
%Book Author, \textit{Book title} (Publisher, place, year) page numbers
% etc
%\end{thebibliography}

\end{document}